\documentclass[11pt]{article}

\usepackage{latexsym,amssymb}

\newtheorem{theorem}{Theorem}
\newtheorem{definition}[theorem]{Definition}
\newtheorem{lemma}[theorem]{Lemma}

\newtheorem{example}[theorem]{Example}

\newcommand{\R}[0]{{\mathbb{R}}}

\newcommand{\Z}[0]{{\mathbb{Z}}}

\def\C{{\mathbb{C}}}
\def\Z{{\mathbb{Z}}}
\def\R{{\mathbb{R}}}

\def\N{{\mathbb{N}}}
\newcommand{\F}[0]{{\mathbb{F}}}

\newcommand{\cH}{{\cal H}}
\newcommand{\cE}{{\cal E}}

\newcommand{\GL}[0]{{\rm GL}}

\newcommand{\onemat}[0]{{\mathbf 1}}
\newcommand{\zeromat}[0]{{0}}

\newcommand{\proof}[1]{\medskip \noindent {\em Proof.} #1 \hfill $\Box$\vspace{0.5cm}}

\newcommand{\ket}[1]{|#1\rangle}
\newcommand{\bra}[1]{\langle #1|}

\title{\Large \textbf{Equivalence of Decoupling Schemes 
and Orthogonal Arrays}} 

\author{Martin R{\"o}tteler\\
Department of Combinatorics and Optimization\\
and Institute for Quantum Computing\\
University of Waterloo\\ Waterloo, Ontario, Canada, N2L 3G1\\
\texttt{mroetteler\symbol{64}math.uwaterloo.ca}\\[2ex]
\and
Pawel Wocjan\\
Institute for Quantum Information\\
California Institute of Technology\\
Pasadena, California 91722, USA\\
\texttt{wocjan\symbol{64}cs.caltech.edu}
}

\date{September 22, 2004}

\begin{document}

\maketitle

\begin{abstract}
We consider the problem of switching off unwanted interactions in a
given multi-partite Hamiltonian. This is known to be an important
primitive in quantum information processing and several schemes have
been presented in the literature to achieve this task. A method to
construct decoupling schemes for quantum systems of pairwise
interacting qubits was introduced by M.~Stollsteimer and G.~Mahler and
is based on orthogonal arrays. Another approach based on triples of
Hadamard matrices that are closed under pointwise multiplication was
proposed by D.~Leung. In this paper, we show that both methods lead to
the same class of decoupling schemes. Moreover, we establish a
characterization of orthogonal arrays by showing that they are
equivalent to decoupling schemes which allow a refinement into
equidistant time-slots. Furthermore, we show that decoupling schemes
for networks of higher-dimensional quantum systems with $t$-local
Hamiltonians can be constructed from classical error-correcting codes.
\end{abstract}

%
%

\section{Introduction}

An important task in the study of quantum systems is to manipulate a
given system Hamiltonian by applying external control operations in
such a way that in effect the time-evolution of some other desired
target Hamiltonian is simulated. Typically, the available control
operations are restricted and furthermore the control schemes employed
are required to be efficient. Hence, the number of control operations
should be a polynomial function in the number of particles which are
governed the system Hamiltonian. In the context of pair-interaction
(also called two-local or two-body) Hamiltonians acting on $n$ qudits
a repertoire of techniques has been developed to use any entangling
Hamiltonian for universal simulation of arbitrary couplings
\cite{knill,BCLLLPV:2002,leung,reversal,finite,NBDCD:2002,DNBT:2002,CLV:2004}.
Here the external control operations available are given by strong
pulses which are local unitary rotations applied to the individual
nodes. A cornerstone of this theory is the development of {\em
decoupling schemes} and {\em selective coupling schemes}. Both are
pulse sequences that switch off all unwanted interactions in a given
Hamiltonian. In the case of a decoupling scheme all interactions have
to be switched off. In contrast the requirement for a selective
coupling scheme is that all interactions except for the interaction
between two fixed nodes have to be switched off.  Two methods have
been proposed to achieve decoupling and selective coupling of a
general pair-interaction Hamiltonians in quantum systems consisting of
$n$ qubits:

\paragraph{Construction I} This method, which was proposed by D.~Leung
\cite{reversal}, uses triples $S_x$, $S_y$, $S_z$ of Hadamard
matrices. If the rows of these matrices satisfy a compatibility
condition, a sequence of pulses around the $\sigma_x$, $\sigma_y$, and
$\sigma_z$-axes in the Bloch sphere can be constructed.

\paragraph{Construction II} In the method put forward by
M.~Stollsteimer and G.~Mahler \cite{stoll} the pulses are constructed
by using orthogonal arrays which are matrices which fulfill a
balancedness conditions between the rows.

\bigskip
The purpose of this paper is to show that these constructions are
equivalent, i.\,e., that each admissible triple of Hadamard matrices
used in Construction I leads to an orthogonal array which can be used
for Construction II (and vice versa). We first show this
correspondence for systems consisting of $n$ two-dimensional systems
and generalize this in the sequel to higher-dimensional systems. Also
the requirement that the given Hamiltonian to be a two-body
Hamiltonian can be relaxed: we show that orthogonal arrays of strength
$t$ can be used to decouple any $t$-local Hamiltonian.

%
%

\section{The Framework: Average Hamiltonian Theory}

Switching off unwanted interactions is an important primitive in the
approaches to render a given Hamiltonian to simulate any other
Hamiltonian
\cite{knill,BCLLLPV:2002,leung,stoll,reversal,finite,DNBT:2002}. Here
simulation is usually understood in a narrow sense in which the
desired target Hamiltonian is approximated up to terms of quadratic
and higher orders. In the following we briefly introduce the facts of
this framework of average Hamiltonian theory \cite{EBW:87,Slichter:90}
which will be needed to develop the theory of decoupling schemes.

Assume that the system Hamiltonian acts on an $n$-fold tensor product
Hilbert space $\cH := \C^d\otimes\C^d\otimes\dots\otimes\C^d$, where
each $\C^d$ denotes the Hilbert space of a so-called {\em qudit}.  Let
$B:=\{\sigma_\alpha\mid \alpha=1,\ldots, d^2-1\}$ be a basis of
traceless matrices acting on $\C^d$. The most general $t$-local
Hamiltonian for a system of $n$ coupled qudits is given by
\begin{equation}\label{eq:pairinteractions}
H:= 
\sum_{s=1}^t \sum_{(k_1,\ldots,k_s)}
\sum_{\alpha_1,\ldots,\alpha_s=1}^{d^2-1}
J_{(k_1,\ldots,k_s);\alpha_1,\ldots,\alpha_s} \,
\sigma_\alpha^{(k_1)} \cdots \sigma_\alpha^{(k_s)}\,,
\end{equation}
where the second sum runs over all $s$-tuples with (different) entries
from $\{1,\ldots,n\}$ and $J_{k_1,\ldots,k_s; \alpha_1, \ldots,
\alpha_s}\in\C$. Here and in the following we use $A^{(k)}$ to denote
the operator that acts as $A$ on the $k$th qubit, i.\,e.,
$A^{(k)}:=\onemat\otimes\cdots\otimes\onemat\otimes A
\otimes\onemat\otimes\cdots\otimes\onemat$.

In the setting discussed in this paper the only possibilities of
external control are given by local unitaries on each qudit. We assume
that it is possible to implement them independently. Formally, all
control operations are elements of some finite subset ${\cal C}$ of
the group ${\cal U}(d)^{\otimes n}$, where ${\cal U}(d)$ denotes the
group of unitary matrices acting on $\C^d$. A useful approximation is
to assume that all operations in ${\mathcal C}$ can be implemented
arbitrarily fast (``fast control limit''). The simulation of
Hamiltonians is based on the following average Hamiltonian
\cite{EBW:87} approach. Let $t_1,t_2,\dots,t_n$ be real numbers and
$V_1,V_2,\dots,V_N \in {\cal C}$ be control operations. Note that
letting the system evolve for some time $t$ (in the following referred
to as ``wait'') has the effect to apply the unitary operator $\exp(- i
H t)$. Hence the sequence
\begin{center}
perform $V_1$, $\;$ wait $t_1$, $\;$ perform $V_2$, $\;$ wait $t_2$,
$\;$ $\ldots$ $\;$ perform $V_N$, $\;$ wait $t_N$
\end{center}
implements the evolution $ \prod_{j=1}^N \exp (-U_j^\dagger H U_j\,
t_j )$, where $U_j=\prod_{i=1}^j V_j$. We say that the scheme consists
of $N$ intervals---sometimes also referred to as {\em
time-slots}---and use the shorthand notation $(t_1, U_1; t_2, U_2;
\ldots; t_N, U_N)$, where we tacitly assume that the underlying
Hamiltonian $H$ is fixed. If the times $t_j$ are small compared to the
time scale of the natural evolution according to $H$ this gives an
approximation to the average Hamiltonian
\[
\bar{H}:=\sum_{j=1}^N t_j U_j^\dagger H U_j / \tau \,,
\]
where $\tau:=\sum_j t_j$ is the slow down factor, i.\,e., the relative
running time of the evolution.

We next introduce decoupling schemes which can be used to simulate the
zero Hamiltonian. For this reason they are also used in dynamical
suppression of decoherence in open quantum systems (``bang-bang''
control), see \cite{bangbang,violaDecoupling,violaControl}.  A
decoupling scheme is a sequence of control operations such that the
resulting average Hamiltonian is the zero matrix for all system
Hamiltonians of the form in eq.~(\ref{eq:pairinteractions}). 

Recall that a unitary operator basis (also called unitary error basis
\cite{knill1}) is a collection of $d^2$ unitaries $U_i$ that are
orthogonal with respect to the inner product $\langle A | B \rangle :=
1/d \, {\rm tr}(A^\dagger B)$. Bases of unitaries with this property
were already studied by Schwinger \cite{Schwinger:60} and recently
several explicit constructions have been found
\cite{knill1,Werner:2001,KR:2003b}.

\begin{definition}[Decoupling scheme]\label{def:decoupling}
A decoupling scheme $D:=(p_1,U_1;\ldots;p_N,U_N)$ is given by
positive real numbers $p_j$ summing up to $1$ and control operations
$U_j\in {\cal C}$ such that
\begin{equation}
\sum_{j=1}^N p_j U_j^\dagger H U_j = \zeromat
\end{equation}
for all $t$-body Hamiltonians acting on $n$ qudits. We call a
decoupling scheme $D$ {\em regular} if the lengths of the time-slots
are the same, i.\,e., if $p_1=p_2=\ldots=p_N$ and in addition if the
operators applied to each node form a unitary error basis.
\end{definition}

Note that if the system consists of one $d$-dimensional node only, a
decoupling scheme is equivalent to a unitary operator basis. This
definition includes decoupling schemes consisting of time-slots of
different length. Many of the decoupling schemes considered in the
literature use only time-slots of equal length
\cite{knill,leung,stoll,reversal,finite}. However, also decoupling
schemes are used in which the intervals have different lengths, most
notably the famous WaHuHa sequence \cite{WHH:68,HW:68,EBW:87}.

The difficulty for systems consisting of more than one node is that we
still want to use a number of operations which is polynomial in the
dimension $d$ of the individual nodes as well as in the number $n$ of
nodes. To construct schemes with this property it is necessary to be
able to apply selective pulses to the nodes \cite{stoll}. In the
following we present two constructions of schemes which achieve
decoupling for any pair of nodes, i.\,e, these schemes can be used to
decouple any pair-interaction Hamiltonian. By applying the same
sequence of pulses to a fixed pair of nodes all interactions will be
switched off with the exception for the bipartite system consisting of
these two nodes. This in turn can be used for universal simulation.

%
%

\section{Hadamard Matrices, Sign Matrices, and Phase Matrices}

In the following we give a short account of the combinatorial objects
underlying the construction used in \cite{reversal} to obtain
decoupling schemes for pair-interaction Hamiltonians acting on $n$
qubits. The construction relies on the concept of so-called sign
matrices which generalize the refocusing schemes for spin echo
experiments on $n$ qubits. The latter have been proposed in
\cite{knill} and are based on Hadamard matrices. Since we will need
Hadamard matrices for the subsequent constructions we briefly recall
their definition.

\paragraph{Hadamard matrices}

We denote the transposed of a matrix $A$ by $A^t$. A Hadamard matrix
of order $N$ is a $\pm 1$ matrix $H_N$ of size $N\times N$ with the
property that $H_N H_N^t = N \onemat_N$. Hadamard matrices have been
studied in combinatorics for a long time and several constructions
have been found. We refer to \cite{BJL:99I,Stinson:2003,CRC} for
background on and constructions of Hadamard matrices. We give some
examples for Hadamard matrices of small order (here and in the
following the entries $\pm 1$ have been abbreviated to $+/-$):
\[ 
H_2 = \left( \begin{array}{cc} + & + \\ + & - \end{array} \right), \quad
H_2 \otimes H_2 = 
\left( \begin{array}{cccc}
++++ \\
+-+- \\
++-- \\
+--+ 
\end{array} \right), \quad A = 
\left( \begin{array}{cccc}
-+++ \\
+-++ \\
++-+ \\
+++- 
\end{array} \right).
\]
It is known that a necessary condition for the existence of a Hadamard
matrix is that either $N=2$ or $N\equiv 0 \; {\rm mod}\; 4$. A long-standing
conjecture is whether indeed for any $N\equiv 0 \; {\rm mod} \; 4$ a Hadamard
matrix of order $N$ exists \cite{BJL:99I}. Since $H_{2^n} := H_2
\otimes \ldots \otimes H_2$ ($n$ tensor factors) is a Hadamard
matrix, we obtain that in dimension $N = 2^n$ at least one Hadamard
matrix exists. 

\paragraph{Sign matrices}

A sign matrix $S_{n,N}$ of size $n \times N$ is given by the first $n$
rows of a Hadamard matrix of order $N$. Hence $S_{n,N}$ is a $\pm 1$
matrix which satisfies $S_{n,N} S_{n,N}^t = N \onemat_n$. Recall that
the Schur product of two $n\times N$ matrices $A$ and $B$ is denoted
by $C := A\circ B$ and is defined as the entry-wise product: $C_{i,j}
:= A_{i,j} B_{i,j}$. As an example we define the following three sign
matrices $S_x$, $S_y$, $S_z$ of size $7 \times 8$:
\[
\setlength{\arraycolsep}{1pt}
S_x := 
\left( \begin{array}{cccccccc}
+ & - & + & - & + & - & + & -  \\
+ & + & - & - & + & + & - & -  \\
+ & - & - & + & + & - & - & +  \\
+ & + & + & + & - & - & - & -  \\
+ & - & + & - & - & + & - & +  \\
+ & + & - & - & - & - & + & +  \\
+ & - & - & + & - & + & + & - 
\end{array}
\right), \; 
S_y := 
\left( \begin{array}{cccccccc}
+ & - & + & - & - & + & - & +  \\
+ & - & + & - & + & - & + & -  \\
+ & + & + & + & - & - & - & -  \\
+ & + & - & - & + & + & - & -  \\
+ & - & - & + & - & + & + & -  \\
+ & - & + & - & - & + & - & +  \\
+ & + & - & - & - & - & + & +  
\end{array}
\right), \;
S_z := 
\left( \begin{array}{cccccccc}
+ & + & + & + & - & - & - & -  \\
+ & - & - & + & + & - & - & +  \\
+ & - & - & + & - & + & + & -  \\
+ & + & - & - & - & - & + & +  \\
+ & + & - & - & + & + & - & -  \\
+ & - & + & - & - & + & - & +  \\
+ & - & + & - & + & - & + & -  
\end{array}
\right).
\]
Besides the fact that they are closed under Schur product, i.\,e.,
$S_x \circ S_y = S_z$, these matrices have another remarkable feature:
all of their rows are actually rows of $H_2 \otimes H_2 \otimes
H_2$. This guarantees that $S_x$, $S_y$, and $S_z$ are sign matrices of
size $7 \times 8$. As we shall see in the next section when we study
criteria for decoupling, these matrices cannot be used to decouple a
general pair-interaction Hamiltonian on seven qubits since all rows of
all three matrices together are not orthogonal (indeed, any row appears
in each of the three matrices). However, if the Hamiltonian of a
seven qubit network is of the particular form where only
$\sigma^{(k)}_x \otimes \sigma^{(\ell)}_x$, $\sigma^{(k)}_y \otimes
\sigma^{(\ell)}_y$, and $\sigma^{(k)}_z \otimes \sigma^{(\ell)}_z$
interaction terms occur in eq.~(\ref{eq:pairinteractions}), these
matrices can be used for decoupling and selective coupling. 

In Section \ref{constrDecoupling} we will present criteria for
decoupling pair-interaction Hamiltonians acting on qubits and show
that they are fulfilled in case we can find sign matrices $S_x$,
$S_y$, and $S_z$ which are closed under taking the Schur product and
have the additional property that their rows are pairwise
orthogonal. The approach \cite{reversal} requires such triples $S_x$,
$S_y$, and $S_z$ of sign matrices of size $n\times N$ which are
related by $S_x \circ S_y = S_z$.

\paragraph{Phase matrices}

The restriction to consider orthogonal matrices with entries $\pm 1$
can be relaxed by allowing the entries to be more general complex
phases. This gives additional flexibility for the decoupling of
pair-interaction Hamiltonians acting on higher dimensional systems
(qudits) and leads to the concept of phase matrices which are defined
as follows: Let $k\in \N$ and let $\omega = \exp(2\pi i/k) \in \C$ be
a primitive $k$th root of unity. Then a phase matrix $P_{n,N}$ of
order $k$ is an $n \times N$ matrix with entries in $\{1, \omega,
\ldots, \omega^{k-1}\}$ such that $P_{n,N} P_{n,N}^\dagger = N
\onemat_n$. Like in case of sign matrices, we are interested in
collections of phase matrices which satisfy certain compatibility
conditions. These conditions can be conveniently stated in terms of
characters of some finite abelian group $G$. For the necessary
background on characters of abelian groups see Appendix
\ref{characters}. In the following we assume that the elements of $G$
are given in a fixed order $g_1, \ldots, g_{|G|}$ and that the
irreducible characters of $G$ are in one-to-one correspondence with
the elements of $G$ and are given by $\{\chi_g : g \in G \}$,
cf.~Theorem \ref{chgroup}.  Recall that the exponent $e(G)$ of $G$ is
the smallest positive integer such that $g^{e(G)} = 1$ for all $g \in
G$. Now, let $P_1, \ldots, P_{|G|}$ be phase matrices of order $e(G)$
which are labeled by the elements of $G$. We say that the $P_g$, where
$g\in G$, are {\em compatible with respect to the Schur product} if
\begin{equation}\label{compatible}
P_g \circ P_h = P_{gh}
\end{equation}
holds for all pairs $g,h \in G$. Note that according to this
definition sign matrices are a special case of phase matrices. Indeed,
we obtain that any sign matrix is a phase matrix for the group $G =
\Z_2 \times \Z_2$ where the phase matrix corresponding to the identity
is given by the all-ones matrix.

The connection to decoupling schemes is as follows: the phase matrix
$P_h$ will describe the phase factors which are acquired when a fixed
unitary matrix $U_h$ is conjugated by some other matrices. The
condition in eq.~(\ref{compatible}) ensures that each vector $v_{k,j}
:= [P_{h; k,j}]_{h \in G}$ for fixed $k = 1, \ldots, n$ and $j = 1,
\ldots, N$ is a homomorphism from $G$ to $\C^\times$. Since the
characters form a group and the order of the elements is fixed we
obtain that $v_{k,j}$ is a row from the character table of $G$,
i.\,e., $v_{k,j} = \chi_g$ for an element $g\in G$. Each element $g\in
G$ corresponds to a control operation $U_g$. These facts will be used
in Section~\ref{schur} to show the vanishing of terms in a
pair-interaction Hamiltonian in case a decoupling scheme is applied.

For general $G$ it turns out to be a non-trivial task to construct
$|G|$-tuples of phase matrices which at the same time fulfill
condition (\ref{compatible}). In Section
\ref{sec:phaseMatricesfromOAs} we will give a construction which is
possible in case the dimension of the nodes is a prime power, i.\,e.,
$|G|=p^m$, where $p$ is prime and $m \in \N$.

%
%

\section{Constructing Decoupling Schemes}
\label{constrDecoupling}

We continue the investigation of decoupling schemes with an
observation concerning the relative lengths of the time-slots in the
scheme. Recall that according to Definition \ref{def:decoupling} a
scheme is regular if all intervals are of equal lengths. In Section
\ref{nonreg} we show that not all decoupling schemes are regular, and
that this is the case even if we are allowed to reorder and refine the
time intervals. In Sections \ref{schur} and \ref{oa} we will then
introduce the two constructions for schemes for decoupling and
selective coupling mentioned in the introduction.

\subsection{Decoupling schemes which are not regular}\label{nonreg}

Assume that a decoupling scheme $D = (p_1, U_1, p_2, U_2, \ldots, p_N,
U_N)$ on a system consisting of $n$ nodes is given. Since $U_j$ is a
local operation for each $j=1, \ldots, N$ we have that $U_j =
U_j^{(1)} \otimes \ldots \otimes U_j^{(n)}$. Focusing on the first two
nodes only, we can always obtain a new decoupling scheme, which has
the form

\bigskip

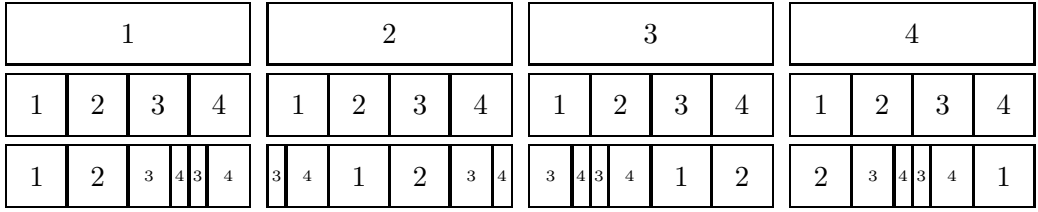
\begin{figure}
\centerline{\unitlength2.3pt
\begin{picture}(43,10) 
\put(0,0){\framebox(40,10){$1$}}
\end{picture}%
\begin{picture}(43,10) 
\put(0,0){\framebox(40,10){$2$}}
\end{picture}%
\begin{picture}(43,10) 
\put(0,0){\framebox(40,10){$3$}}
\end{picture}%
\begin{picture}(43,10) 
\put(0,0){\framebox(40,10){$4$}}
\end{picture}%
}%
\smallskip
\centerline{\unitlength2.3pt%
\begin{picture}(43,10) 
\put(0,0){\framebox(10,10){$1$}}
\put(10,0){\framebox(10,10){$2$}}
\put(20,0){\framebox(10,10){$3$}}
\put(30,0){\framebox(10,10){$4$}}
\end{picture}%
\begin{picture}(43,10) 
\put(0,0){\framebox(10,10){$1$}}
\put(10,0){\framebox(10,10){$2$}}
\put(20,0){\framebox(10,10){$3$}}
\put(30,0){\framebox(10,10){$4$}}
\end{picture}%
\begin{picture}(43,10) 
\put(0,0){\framebox(10,10){$1$}}
\put(10,0){\framebox(10,10){$2$}}
\put(20,0){\framebox(10,10){$3$}}
\put(30,0){\framebox(10,10){$4$}}
\end{picture}%
\begin{picture}(43,10) 
\put(0,0){\framebox(10,10){$1$}}
\put(10,0){\framebox(10,10){$2$}}
\put(20,0){\framebox(10,10){$3$}}
\put(30,0){\framebox(10,10){$4$}}
\end{picture}%
}%
\smallskip
\centerline{\unitlength2.3pt%
\begin{picture}(43,10) 
\put(0,0){\framebox(10,10){$1$}}
\put(10,0){\framebox(10,10){$2$}}
\put(20,0){\framebox(7,10){\tiny $3$}}
\put(27,0){\framebox(3,10){\tiny $4$}}
\put(30,0){\framebox(3,10){\tiny $3$}}
\put(33,0){\framebox(7,10){\tiny $4$}}
\end{picture}%
\begin{picture}(43,10) 
\put(0,0){\framebox(3,10){\tiny $3$}}
\put(3,0){\framebox(7,10){\tiny $4$}}
\put(10,0){\framebox(10,10){$1$}}
\put(20,0){\framebox(10,10){$2$}}
\put(30,0){\framebox(7,10){\tiny $3$}}
\put(37,0){\framebox(3,10){\tiny $4$}}
\end{picture}%
\begin{picture}(43,10) 
\put(0,0){\framebox(7,10){\tiny $3$}}
\put(7,0){\framebox(3,10){\tiny $4$}}
\put(10,0){\framebox(3,10){\tiny $3$}}
\put(13,0){\framebox(7,10){\tiny $4$}}
\put(20,0){\framebox(10,10){$1$}}
\put(30,0){\framebox(10,10){$2$}}
\end{picture}%
\begin{picture}(43,10) 
\put(0,0){\framebox(10,10){$2$}}
\put(10,0){\framebox(7,10){\tiny $3$}}
\put(17,0){\framebox(3,10){\tiny $4$}}
\put(20,0){\framebox(3,10){\tiny $3$}}
\put(23,0){\framebox(7,10){\tiny $4$}}
\put(30,0){\framebox(10,10){$1$}}
\end{picture}%
}
\caption[]{\label{nonregular} A decoupling scheme for a system of
three qubits which is not regular, i.\,e., the time-slots cannot be
rearranged into a form where all time-slots have the same length. The
transformations applied to the individual qubits correspond to the
Pauli matrices as follows: $1=\onemat_2$, $2=\sigma_x$, $3=\sigma_y$,
and $4=\sigma_z$. The time intervals indicated in the figure have four
different basic lengths $t_1$ , $t_2 = 1/4 \, t_1$,
$t_3=\frac{1}{\sqrt{2}} \, t_2$, and $t_4 = \frac{\sqrt{2}-1}{\sqrt{2}}\,
t_2$. For instance, the first intervals applied to the first qubit all
length $t_1$, whereas the interval lengths for qubit three are given
by $t_2$, $t_2$, $t_3$, $t_4$, $t_4$, $t_3$, etc.}
\end{figure}

\begin{tabular}{cccc}
\framebox[4.068cm]{$r U_1^{(1)}$} &
\framebox[4.068cm]{$r U_2^{(1)}$} & \ldots & 
\framebox[4.068cm]{$r U_{N}^{(1)}$} \\[1ex]
\framebox[1.5cm]{$s U_1^{(2)}$} \, \ldots \, \framebox[1.5cm]{$s U_{N}^{(2)}$} &
\framebox[1.5cm]{$s U_1^{(2)}$} \, \ldots \, \framebox[1.5cm]{$s U_{N}^{(2)}$} &
\ldots &
\framebox[1.5cm]{$s U_1^{(2)}$} \, \ldots \, \framebox[1.5cm]{$s U_{N}^{(2)}$} 
\end{tabular},

\bigskip

\noindent
where $r,s\in \R$ are such that $r N = s N^2 = \sum_{j=1}^N {p_j} =
1$. However, in order to obtain a regular scheme in general we cannot
continue in this fashion to more than three nodes.  The pulse sequence
given in Figure \ref{nonregular} provides an example of a decoupling
scheme that cannot be refined into time-slots which have the same
lengths. The control operations used in the scheme are the Pauli
matrices, which form a basis for the vector space of all $2 \times 2$
matrices and are given by
\[
\onemat_2 = \left(\begin{array}{rr}
1 & 0\\ 0 & 1 
\end{array}\right), \quad
\sigma_x = \left(\begin{array}{rr}
0 & 1\\ 1 & 0 
\end{array}\right), \quad
\sigma_y = \left(\begin{array}{rr}
0 & -i\\ i & 0 
\end{array}\right), \quad
\sigma_z = \left(\begin{array}{rr}
1 & 0\\ 0 & -1 
\end{array}\right).
\]
In the given example the network consists of three qubits and in each
time-slot precisely one of the four Pauli matrices is applied.
Indeed, we first verify that the pulse sequence defined in Figure
\ref{nonregular} defines a decoupling scheme for any pair-interaction
Hamiltonian: first, note that the sum of the times for each Pauli
operator applied to the individual qubits is constant, i.\,e., the
local terms are removed. Moreover, by considering pairs of rows we
verify directly that also any pair of symbols $(a,b)$ with $a,b \in
\{1,2,3,4\}$ is applied for the same time $t_2$. For example in case
of rows two and three we obtain for the pairs $(1,1)$ the total time
$t_2$ and for $(3,4)$ the total time $t_3 + t_4 = t_2$. However, the
sequence cannot be subdivided into a finite number of intervals of
equal lengths. Indeed, this would contradict the fact that $\sqrt{2}$
is not a rational number.

%
%

\subsection{Decoupling schemes from sign and phase matrices}\label{schur}

We now describe the approach of \cite{reversal} to construct
decoupling and selective coupling schemes for general pair-interaction
Hamiltonians acting on qubits. Then we present a generalization for
qudits by generalizing the underlying group-theoretical structures
based on our definition of phase matrices.

\paragraph{The qubit case: decoupling schemes from sign matrices}

A general pair-interaction Hamiltonian, i.\,e., two-local
Hamiltonian, for $n$ qubits may be written in the form
\begin{equation}\label{eq:pairinteractions2}
H:=\sum_{k<\ell} \sum_{\alpha\beta} 
J_{k\ell;\alpha\beta} \sigma_\alpha^{(k)} \sigma_\beta^{(\ell)}
+
\sum_k \sum_\alpha J_{k;\alpha} \sigma_\alpha^{(k)}
\end{equation}
where $J_{k\ell;\alpha\beta}\in\R$, $J_{k;\alpha}\in\R$ and where
$\sigma_\alpha$ are the Pauli matrices, i.\,e., $\alpha \in\{x,y,z\}$.
To construct decoupling schemes we choose $\onemat_2$, $\sigma_x$,
$\sigma_y$ and $\sigma_z$ as control operations.  Then in each
time-slot of a decoupling scheme each of the $n$ qubits is conjugated
by precisely one of the matrices $\onemat_2$, $\sigma_x$, $\sigma_y$,
and $\sigma_z$. Since the terms of the Hamiltonian in
eq.~(\ref{eq:pairinteractions2}) are expressed in terms of the Pauli
matrices we have to compute the resulting effect by this local
conjugation: the possible sign assignments for
$\onemat_2^{(k)},\sigma_x^{(k)},\sigma_y^{(k)},\sigma_z^{(k)}$ are
given by the following table:

\begin{equation}\label{eq:signs}
\begin{array}{|c||cccc|}
\hline
         & \onemat_2 & \sigma_x & \sigma_y & \sigma_z \\ \hline\hline
\onemat_2  & + & + & + & + \\
\sigma_x & + & + & - & - \\
\sigma_y & + & - & + & - \\
\sigma_z & + & - & - & +  \\
\hline
\end{array}
\end{equation}

\medskip
\noindent Here the rows are labeled by the operators used in
eq.~(\ref{eq:pairinteractions2}) and the columns by the operators
realizing the conjugation at each time-slot. Note that in each column
the signs multiply to $+1$ since conjugation by a local unitary
corresponds to a $SO(3)$ rotation of the Bloch vector. Hence, the
signs assignments of the three Pauli matrices $\sigma_\alpha^{(k)}$
acting on the same qubit $k$ are not independent but rather the third
is always given by the other two.

Now, in each interval $\sigma_\alpha^{(k)}$ acquires either a $+$ or a
$-$ sign, which is controlled by the applied local unitaries (the
identity matrix or a Pauli matrix). According to table
(\ref{eq:signs}) the bilinear coupling
$J_{k\ell;\alpha\beta}\sigma_\alpha^{(k)}\sigma_\beta^{(\ell)}$ is
unchanged (negated) when the signs of $\sigma_\alpha^{(k)}$ or
$\sigma_\alpha^{(\ell)}$ agree (disagree).

\paragraph{Decoupling criteria in terms of sign matrices}

We show that sign matrices satisfying certain orthogonality conditions
yield decoupling schemes. The $(k,j)$ entry of $S_\alpha$ for $\alpha
\in \{x,y,z\}$ is denoted by $S_{\alpha;kj}$ and gives the sign of
$\sigma_\alpha^{(k)}$ in the $j$th time-slot. Hence a regular
decoupling scheme which uses $N$ time-slots can be obtained from these
matrices using the following rules: if the triple of
entries at position $(k,j)$ is given by $(+++)$,$(+--)$,$(-+-)$,
respectively $(--+)$ then the operation applied to qubit $k$ in time
step $j$ is given by $\onemat_2$, $\sigma_x$, $\sigma_y$, respectively
$\sigma_z$.

Decoupling is achieved if any two rows taken from $S_x$, $S_y$, $S_z$
are orthogonal. To achieve selective coupling between two nodes $k$
and $\ell$, the operations applied to the nodes $k$ and $\ell$ are
chosen to be identical while still maintaining orthogonality of the
modified sign matrices \cite{reversal}. Within this framework
sufficient and necessary conditions for decoupling of all
pair-interactions are given by the following equations:
\begin{equation}
\sum_{j=1}^N S_{\alpha;kj}=0
\end{equation}
for all $\alpha$ and all $k$, and
\begin{equation}
\sum_{j=1}^N S_{\alpha;kj} S_{\beta;\ell j}=0
\end{equation}
for all $\alpha,\beta$ and all $k<\ell$. The first condition ensures
that all local terms are removed and the second condition that all
bilinear terms are removed. These conditions are satisfied if the sign
matrices $S_x$, $S_y$, $S_z$ and all rows of all three matrices are
orthogonal to each other.

\begin{example}\rm

As an example consider the following sign matrices which specify a
decoupling scheme for a system of $5$ qubits with $16$ time-slots.
\begin{equation}\label{signsx}
S_x := \left(
\begin{array}{cccccccccccccccc}
+ & + & + & + & + & + & + & + & - & - & - & - & - & - & - & - \\
+ & + & - & - & - & - & + & + & + & + & - & - & - & - & + & + \\
+ & - & - & + & + & - & - & + & - & + & + & - & - & + & + & - \\
+ & - & - & + & - & + & + & - & - & + & + & - & + & - & - & + \\
+ & + & - & - & + & + & - & - & - & - & + & + & - & - & + & + 
\end{array}
\right),
\end{equation}
\begin{equation}\label{signsy}
S_y := 
\left(
\begin{array}{cccccccccccccccc}
+ & + & + & + & - & - & - & - & + & + & + & + & - & - & - & - \\
+ & - & + & - & - & + & - & + & + & - & + & - & - & + & - & + \\
+ & + & - & - & + & + & - & - & + & + & - & - & + & + & - & - \\
+ & + & - & - & - & - & + & + & - & - & + & + & + & + & - & - \\
+ & - & + & - & - & + & - & + & - & + & - & + & + & - & + & - 
\end{array}
\right),
\end{equation}
\begin{equation}\label{signsz}
S_z := 
\left(
\begin{array}{cccccccccccccccc}
+ & + & + & + & - & - & - & - & - & - & - & - & + & + & + & + \\
+ & - & - & + & + & - & - & + & + & - & - & + & + & - & - & + \\
+ & - & + & - & + & - & + & - & - & + & - & + & - & + & - & + \\
+ & - & + & - & + & - & + & - & + & - & + & - & + & - & + & - \\
+ & - & - & + & - & + & + & - & + & - & - & + & - & + & + & - 
\end{array}
\right).
\end{equation}
In general the construction of admissible triples $(S_x, S_y, S_z)$ of
sign matrices proves to be a delicate task. However, a construction of
triples of sign matrices of size $(2^{2n}-1)/3 \times 2^{2n}$ by
partitioning the rows of the Hadamard matrices $H_2^{\otimes 2n}$,
where $n\in\N$, was given in \cite{reversal}. This construction has
been revisited in \cite{finite} where an alternative proof based on
spreads in a finite geometry has been given. We will give yet another
proof of this family of sign matrices in
Section~\ref{sec:phaseMatricesfromOAs} which is based on Hamming
codes.

\end{example}

\paragraph{Generalization to the qudit case: phase matrices}

In the following we generalize the approach described in
\cite{reversal} to pair-interactions between higher-dimensional
systems, i.\,e., qudits. It will be useful to express a general
pair-interaction Hamiltonian with respect to a so-called nice error
basis and to use the matrices from such basis as control
operations. First, we recall the definition of nice error bases
\cite{knill1,knill2,KR:2002}.

\begin{definition}[Nice error basis]
Let $G$ be a group of order $d^2$ with identity element $e$. A
\emph{nice error basis} on $\mathbb{C}^d$ is a set
$\mathcal{E}=\{U_g\in\C^{d\times d}\mid g\in G\}$ of unitary matrices,
which are labeled by the elements of $G$, such that (i) $U_e$ is the
identity matrix, (ii) $\mathrm{tr}\, U_g=d\,\delta_{g,e}$ for all
$g\in G$, and (iii) $U_g U_h=\alpha(g,h)U_{gh}$ for all $g,h\in
G$. The factor system $\alpha(g, h)$ is a function from $G\times G$ to
the set $\C^\times := \C \setminus \{0\}$.
\end{definition}
Condition (ii) shows that the matrices $U_g$ are pairwise orthogonal
with respect to the trace inner product. The group $G$ is called {\em
index group} since its group elements index the elements of the nice
error basis $\cE$.

\begin{example}
Let $d \in \N$ and let $\omega = \exp(2 \pi i/d)$ denote a primitive
$d$-th root of unity. Next, we define operators $S := \sum_{k=0}^{d-1}
\ket{k}\bra{k+1}$, where the indices are reduced modulo $d$, and $T :=
\sum_{k=0}^{d-1} \omega^k \ket{k}\bra{k}$. Then the set ${\cal E}_d :=
\{ S^i T^j : i = 0, \ldots, d-1, j = 0, \ldots, d-1 \}$ is a nice
error basis on $\C^d$ (see, e.\,g., \cite{AK:2001}). This shows the
existence of nice error bases for any dimension $d\in \N$.  In this
case the index group is the abelian group $G = \Z_d \times \Z_d$. The
identity $ST = \omega TS$ is readily verified. This shows that the
corresponding factor system $\alpha$ is given by $\alpha( (i, j), (k,
\ell) ) = \omega^{-jk}$, for all $(i,j), (k,\ell) \in G$.
\end{example}

\noindent Next, we describe a particular way of representing a general
pair-interaction Hamiltonian acting on $n$ qudits. Let ${\cal E} = \{
U_g : g \in G\}$ be a nice error basis. Since the matrices $U_g$ form
a basis of $\C^{d\times d}$ a general pair-interaction Hamiltonian on
$n$ coupled qudits may be written as
\begin{equation}\label{eq:niceHamiltonian}
H:=\sum_{k<\ell} 
\sum_{h,h'\neq e} 
J_{k \ell;h h'} U_h^{(k)} U_{h'}^{(\ell)}
+
\sum_k \sum_{h\neq e} J_{k;h} U_h^{(k)}
\,,
\end{equation}
where the coefficients $J_{k \ell;h h'}\in\C$ and $J_{k;h}\in\C$ are
chosen such that $H$ is a traceless Hermitian matrix. In the following
we consider decoupling schemes with time-slots of equal length and
with elements of a nice error basis having abelian index groups as
control operations. Then, in each time-slot, each $U_h$ in
eq.~(\ref{eq:niceHamiltonian}) acquires a phase factor that is
controlled by the applied local unitaries of the nice error basis. We
define $\chi(g,h)$ to be the phase factor that $U_h$ acquires when it
is conjugated by $U_g$, i.\,e.\ $\chi(g,h)$ is defined via the relation
\begin{equation}\label{eq:phase}
U_g^\dagger U_h U_g = \chi(g,h) U_h\,.
\end{equation}
Hence, the bilinear term $U_h\otimes U_{h'}$ acquires the phase factor
$\chi(g,h)\chi(g',h')$ if it is conjugated by $U_g\otimes U_{g'}$.
The fact that conjugation by $U_g$ of $U_h$ merely introduces a phase
factor follows from the index group being abelian. We need to
characterize the corresponding $d^2\times d^2$ matrix
\begin{equation}\label{eq:phasefactors}
{\cal X}:=\left( \chi(g,h) \right)_{g,h\in G}
\end{equation} 
with entries $\chi(g,h)$.  

\begin{lemma}\label{lemma:character}
Let $\cE:=\{U_g\mid g\in G\}$ be a nice error basis with an abelian
index group $G$. Then the matrix ${\cal X} = \left( \chi(g,h)
\right)_{g,h\in G}$ is the character table of the group $G$.
\end{lemma}

We give a proof in Appendix \ref{appendixB}. This lemma shows that
${\cal X}$ is a character table of $G$ and thereby provides a natural
generalization of Table \ref{eq:signs} to non-qubit systems.  Given a
decoupling scheme which contains operators from a nice error basis
${\cal E} = \{ U_g : g \in G\}$ we obtain a collection of phase
matrices $P_h$ for each $h \in G$ in the following way: the $(k,j)$th
entry of $P_h$---denoted by $P_{h;k,j}$---is given by the phase factor
of $U_h$ that is acquired in the $j$th time-slot by conjugating the
$k$ qudit by the respective element of the scheme.

\paragraph{Criteria for decoupling in term of phase matrices} 

Let $P_h$ be phase matrices. Then they define a decoupling scheme if
and only if
\begin{equation}\label{eq:removeLocal}
\sum_{j=1}^N P_{h;kj}=0
\end{equation}
for all $k$ and for all $h\neq e$, and
\begin{equation}\label{eq:removeBilinear}
\sum_{j=1}^N P_{h;kj} P_{h';\ell j}=0
\end{equation}
for all $k<\ell$ and for all $h,h'\neq e$. The condition given in
eq.~(\ref{eq:removeLocal}) ensures that all local terms are removed.
Eq.~(\ref{eq:removeBilinear}) implies that all bilinear terms are
removed. The question how to construct collections of phase matrices
which satisfy eqs.~(\ref{eq:removeLocal}) and
(\ref{eq:removeBilinear}) is addressed in Section
\ref{sec:phaseMatricesfromOAs} and a solution is presented if all
nodes have a dimension which is a power of a prime.

%
%

\subsection{Decoupling schemes from orthogonal arrays}\label{oa}

Orthogonal arrays have been applied in the design of experiments to
plan statistical data collections systematically. The books
\cite{BJL:99I,CRC,HSS:99} provide good introductions to the topic.  In
the following we recall the definition of orthogonal arrays (or OAs
for short) and show how to use them for the decoupling problem.
M.~Stollsteimer and G.~Mahler have first used OAs for the construction
of decoupling schemes and selective coupling schemes \cite{stoll} for
pair-interaction Hamiltonians acting on qubits. This method was
generalized to pair-interaction Hamiltonians acting on qudit
\cite{finite} with the helps of unitary error bases. Even more
generally, we show in this section how to use orthogonal arrays for
constructing decoupling schemes for $t$-local Hamiltonians acting on
qudits with $t>2$.

\begin{definition}[Orthogonal array of strength $t$]
Let ${\cal A}$ be a finite set and let $n, N \in \N$. An $n\times N$
array $M$ with entries from ${\cal A}$ is an orthogonal array with
$|{\mathcal A}|$ levels, strength $t$, and index $\lambda$ if and only
if every $t \times N$ sub-array of $M$ contains each possible
$t$-tuple of elements in ${\cal A}$ precisely $\lambda$ times as a
column. We use the notation $OA_\lambda(N,n,s,t)$ to denote a
corresponding orthogonal array. If $\lambda$, $s$, and $t$ are
understood we also use the shorthand notation $OA(N,n)$.
\end{definition}

\begin{table}
\begin{center}
\begin{tabular}{|c||c||c|}
\hline
Parameter & Design of Experiments & Decoupling Schemes for qudit systems \\
\hline
\hline
$n$ & factors & nodes (qudits) \\
$N$ & runs & time-slots \\
${\cal A}$ & levels & elements of an operator basis \\
$d$ & number of levels & (dimension of the nodes$)^2$ \\
$t$ & strength & locality \\
$\lambda$ & index & --- \\
\hline
\end{tabular}
\caption{\label{dict} Dictionary between notions used in the theory of
design of experiments and the theory of qudit systems to describe the
parameters of an orthogonal array $OA_\lambda(N,n,d,t)$ over alphabet
${\cal A}$. }
\end{center}
\end{table}

In statistics for the various parameters of an orthogonal array (OA)
some traditional terminology is used. In the context of decoupling and
simulation of Hamiltonians different names for these parameters are
used. We provide a dictionary between the different languages in table
\ref{dict}. Note that as a convention we write OAs as $n \times N$
matrices, whereas most authors in design theory prefer to write the
matrices as $N \times n$ matrices. Besides typographic reasons we
found the presentation using $n\times N$ matrices useful to establish
the correspondence with pulse sequences in NMR which are typically
read from left to right like a musical score \cite{EBW:87}.

An important special case arises if the strength $t$ is two. This
means that each pair of elements of ${\cal A}$ occurs $\lambda$ times
in the list $((a_{kj},a_{\ell j})\mid j=1,\ldots N)$ for $1\le k<\ell\le n$.
Most of the known constructions actually yield arrays of strength two
\cite{HSS:99}. For many physical systems it will be sufficient to
study arrays of small strength since the strength relates to the
degree of the interactions, i.\,e., for pair-interaction Hamiltonians
it is sufficient to consider arrays of strength $t=2$. 

\begin{example}\rm 
As an example of small size we give an orthogonal array with
parameters $OA(16,5,4,2)$. This means that we have $16$
runs/time-slots, $5$ factors/qubits, $4$ different symbols/pulses, and
(pair-interaction) strength two.
\[
\left(
\begin{array}{cccccccccccccccc}
1 & 1 & 1 & 1 & 2 & 2 & 2 & 2 & 3 & 3 & 3 & 3 & 4 & 4 & 4 & 4 \\
1 & 2 & 3 & 4 & 4 & 3 & 2 & 1 & 1 & 2 & 3 & 4 & 4 & 3 & 2 & 1 \\
1 & 3 & 4 & 2 & 1 & 3 & 4 & 2 & 3 & 1 & 2 & 4 & 3 & 1 & 2 & 4 \\
1 & 3 & 4 & 2 & 4 & 2 & 1 & 3 & 4 & 2 & 1 & 3 & 1 & 3 & 4 & 2 \\
1 & 2 & 3 & 4 & 2 & 1 & 4 & 3 & 4 & 3 & 2 & 1 & 3 & 4 & 1 & 2 
\end{array}
\right)
\]
It is straightforward to check that indeed every pair of rows contains
all the $16$ possible pairs of symbols precisely once. This array was
obtained from a linear error-correcting code over the finite field
$\F_4$. We will explore this construction in more detail in
Theorem~\ref{OAcodes}.
\end{example}

Next, we generalize the ideas in \cite{stoll,finite} and describe how
to achieve decoupling of $t$-local Hamiltonian acting on qudits. The
basic idea is to use an orthogonal array $M$ with parameters
$OA(N,n,d^2,t)$ over an alphabet ${\cal A}$ of size $d^2$. Here $d$
denotes the dimension of the qudits. The elements of ${\cal A}$ are
identified with the operators $U_1,\ldots, U_{d^2}$ of a unitary error
basis \cite{finite}. The orthogonal array is an $n\times N$ matrix
$M=(m_{kj})$ and determines which elements of the unitary error basis
are used on the qudits in the time-slots as follows:
\begin{equation}
\frac{1}{N}
\sum_{j=1}^N 
(U_{m_{1j}}^\dagger\otimes\cdots\otimes U_{m_{nj}}^\dagger)
H
(U_{m_{1j}}\otimes\cdots\otimes U_{m_{nj}}).
\end{equation}
The following theorem shows that the resulting time evolution is that
of the zero Hamiltonian which means that indeed the decoupling
conditions given in Definition \ref{def:decoupling} are satisfied.

\begin{theorem}[Decoupling with OAs]
Any orthogonal array $OA(N,n,d^2,t)$ over an alphabet of size $d^2$
can be used to decouple $n$ qudits which are governed by a $t$-local
Hamiltonian within $N$ time-slots.
\end{theorem}
\proof{ First, note that for any $t$-local Hamiltonian $H$ of a system
consisting of $t$ qudits the following operations define a decoupling
scheme \cite{finite}: Let $U_1,U_2,\ldots,U_{d^2}$ denote the elements
of some unitary error basis for $\C^d$. Since the tensor products of
all possible pairs of these elements form a vector space basis of the
linear maps acting on $(\C^d)^{\otimes t}$ we obtain that
\cite{Werner:2001,finite}
\begin{equation}\label{eq:bipartite}
\frac{1}{d^{2t}}\sum_{i_1, \ldots, i_t=1}^{d^2} 
(U_{i_1}^\dagger \otimes \ldots \otimes U_{i_t}^\dagger) 
H 
(U_{i_1} \otimes \ldots \otimes U_{i_t}) = \zeromat
\end{equation}
for all (traceless) Hamiltonians acting on $(\C^d)^{\otimes
t}$. Recall that we assume without loss of generality that $H$ is
traceless. Let $B:=\{\sigma_1, \sigma_2,\ldots,\sigma_{d^2-1}\}$ be a
basis for the vector space of traceless matrices of size $d \times
d$. Recall that a general $t$-local Hamiltonian acting on $n$ qudits
can be written as
\begin{equation}
H=\sum_{s=1}^t\,
\sum_{(k_1, \ldots, k_s)}\, 
\sum_{\alpha_1, \ldots, \alpha_s=1}^{d^2-1} 
J_{(k_1, \ldots, k_s);\alpha_1, \ldots, \alpha_s} \, 
\sigma_{\alpha_1}^{(k_1)}\ldots\sigma_{\alpha_s}^{(k_s)}
\end{equation}
where the second sum runs over all $s$-tuples with different entries
from $\{1,\ldots,n\}$ and $J_{(k_1, \ldots, k_s);\alpha_1, \ldots,
\alpha_s}\in \C$. Now, we pick any $s$-subset $\{k_1, \ldots, k_s\}
\subseteq \{1, \ldots, n\}$ of the nodes and denote by $C_{k_1,
\ldots, k_s}$ the coupling between these nodes. We define
$C_{k_1,\ldots,k_s}$ to be the coupling among the qudits
$k_1,\ldots,k_s$, i.\,e.,
\[
C_{k_1,\ldots,k_s}:=
\sum_{\alpha_1, \ldots, \alpha_s} 
J_{(k_1, \ldots, k_s);\alpha_1, \ldots, \alpha_s} \, 
\sigma_{\alpha_1}^{(k_1)}\otimes \ldots\otimes \sigma_{\alpha_s}^{(k_s)}\,.
\]
We define $\hat{C}_{k_1,\ldots,k_s}$ to be the corresponding operator
acting on $(\C^d)^{\otimes s}$. Formally, we say that
$C_{k_1,\ldots,k_s}$ is obtained by embedding
$\hat{C}_{k_1,\ldots,k_s}$ into $(\C^d)^{\otimes n}$ according to the
tuple $(k_1,\ldots,k_s)$. For any operator $X$ acting on
$(\C^d)^{\otimes s}$ we denote the embedding into $(\C^d)^{\otimes n}$
according to the tuple $(k_1,\ldots,k_s)$ by $X^{(k_1,\ldots,k_s)}$.

The idea of the proof is to reduce the problem to
eq.~(\ref{eq:bipartite}) by using the local structure of the
Hamiltonian. Since $M$ is an $OA(N,n,d^2,t)$ all elements of
$\{1,2,\ldots,d^2\}^s$ for $s\le t$ appear equally often in the list
$(m_{k_1,j},\ldots, m_{k_s,j})$ where $j=1,\ldots,N$. Therefore, the
average Hamiltonian corresponding to the coupling among the qudits
$k_1,\ldots,k_s$ is evaluated as follows:
\begin{eqnarray*}
&&
\frac{1}{N} 
\sum_{j=1}^N 
(U_{m_{1j}}^\dagger\otimes\cdots\otimes U_{m_{nj}}^\dagger) 
\, C_{k_1,\ldots,k_s} \,
(U_{m_{1j}}\otimes\cdots\otimes U_{m_{nj}}) \\
& = &
\left[\frac{1}{N}
\sum_{j=1}^N 
(U_{m_{k_1,j}}^\dagger \otimes\cdots\otimes U_{m_{k_s,j}}^\dagger)
\, \hat{C}_{k_1,\ldots,k_s} \,
(U_{m_{k_1,j}} \otimes \cdots\otimes U_{m_{k_s,j}})
\right]^{(k_1,\ldots,k_s)} \\
& = &
\left[
\frac{1}{d^{2t}}
\sum_{i_1, \ldots, i_s=1}^{d^2}
(U_{i_1}^\dagger
\otimes \ldots \otimes U_{i_s}^\dagger) 
\, \hat{C}_{k_1,\ldots k_s} \, 
(U_{i_1} \otimes \ldots \otimes U_{i_s}) 
\right]^{(k_1,\ldots,k_s)} = 
\zeromat\,.
\end{eqnarray*}
The equality between the second last and last line is due to
eq.~(\ref{eq:bipartite}).}

%
%

\section{Equivalence of the Constructions}
\label{sec:equivalence} 

We show that the methods based on phase matrices and orthogonal arrays
of strength two lead to the same class of decoupling schemes if we use
elements of a nice error basis with an abelian index group as control
operations. More precisely, we prove that the decoupling conditions
given in eqs.~(\ref{eq:removeLocal}) and (\ref{eq:removeBilinear}) are
equivalent to the condition that the decoupling matrix is an
orthogonal array of strength two.

\subsection{Phase matrices from orthogonal arrays}
\label{sec:phaseMatricesfromOAs}

We show that a decoupling scheme based on an orthogonal arrays of
strength two gives rise to phase matrices satisfying the decoupling
criteria eqs.~(\ref{eq:removeLocal}) and (\ref{eq:removeBilinear}) for
pair-interaction (two-local) Hamiltonians.

\begin{theorem}\label{th:phaseMatricesfromOAs}
Let $G$ be a finite abelian group of exponent $e(G)$ and let
$\cE=\{U_g \mid g\in G\}$ be a nice error basis for $\C^d$ with index
group $G$. Then a decoupling scheme for $n$ qudits governed by a
pair-interaction Hamiltonian can be constructed. The scheme uses $N$
time-slots and is constructed from an orthogonal array $OA(N,n,d^2,2)$
over the alphabet ${\mathcal A}=\{1,2,\ldots d^2\}$. Furthermore, the
scheme gives rise to phase matrices $[P_{h}]_{h\in G}$ of size
$n\times N$ with entries in $\{1,\omega,\ldots,\omega^{e(G)-1}\}$ which
satisfy the orthogonality conditions and are compatible with respect
to taking Schur products.
\end{theorem}

\proof{ We denote the orthogonal array by $M=[m_{k,j}]$, where
$k=1,\ldots, n$ and $j=1, \ldots, N$. Next, we fix an ordering
$g_1,\ldots,g_{d^2}$ of the elements of $G$ and assume that $g_1 = e$
is the identity. Moreover, we identify the operators of $\cE$ with the
elements of ${\mathcal A}$ according to $1 \mapsto g_1$, $2 \mapsto
g_2,\ldots, d^2\mapsto g_{d^2}$. Note that conjugating $U_h\in \cE$ by
$U_g\in\cE$ results in a phase factor that is the $(g,h)$th entry of
the character table of $G$.

Starting from the given orthogonal array we construct $d^2$ phase
matrices $P_{g_1},\ldots,P_{g_{d^2}}$ as follows. Pick row number $k$
of the OA and replace each symbol according to $a \mapsto v_{g_a}$ for
$a=1,\ldots d^2$ where $v_{g_a}$ denotes the $g_a$th row of the
character table of $G$ (cf.~Lemma \ref{lemma:character}). By assigning
$P_{g_i}$ for $i=1,\ldots,d^2$ the first, second, and $d^2$th
components of each entry, we define the rows number $k$ of the $d^2$
phase matrices. In other words, the entry $(k,j)$ of the matrix $P_h$,
where $h \in \{g_1, \ldots, g_{d^2}\}$, is given by $P_{h; k,j} =
\chi(m_{k,j}, h)$. Note that the matrix $P_{g_1}$ is the all-ones
matrix of size $n \times N$. 

While the condition $P_g \circ P_h = P_{gh}$ is automatically
guaranteed since the characters form a group, we have to show that the
resulting vectors are pairwise orthogonal. In order to do so we pick
two rows $k$ and $\ell$ of the original orthogonal array. We may
assume that the two rows have the following form (or else we apply a
suitable permutation of the columns)
\begin{equation}\label{normalform}
\Big(
\underbrace{
\begin{array}{cccc|cccc|c|cccc}
1 & 1 & \ldots & 1 & 2 & 2 & \ldots & 2 & \quad\ldots\quad & d^2 & d^2 & 
\ldots & d^2 \\
1 & 2 & \ldots & d^2 & 1 & 2 & \ldots & d^2 & \quad\ldots\quad & 1 & 2 & 
\ldots & d^2 
\end{array}}_{\lambda \mbox{ times}}
\Big)
\end{equation}
since all pairs appear equally often ($\lambda$ times) in the OA.  Let
$\vec{\lambda}=(++\cdots+)$ be the vector of length $\lambda=N/d^4$
containing only the entry $+1$. Furthermore, for each $h \in \{g_1,
\ldots, g_{d^2}\}$ define a vector $w_h \in \C^{d^2}$ as follows. We
define $w_h := [\chi(g_1, h), \ldots, \chi(g_{d^2},h)]$, i.\,e., $w_h$
is the $h$th column of the character table of $G$ (cf.~Lemma
\ref{lemma:character}). By substituting the entries of rows $k$ and
$\ell$ in the form of eq.~(\ref{normalform}) by the corresponding sign
assignments in Table~(\ref{eq:signs}) we obtain as the $k$th rows of
$P_{g_1}, \ldots, P_{g_{d^2}}$ the vectors
\[
\vec{\lambda}\otimes w_{g_1}     \otimes (+\cdots +)\,,\quad
\vec{\lambda}\otimes w_{g_2}     \otimes (+\cdots +)\,,\quad\ldots\,,\quad
\vec{\lambda}\otimes w_{g_{d^2}} \otimes (+\cdots +)
\]
and for the $\ell$th rows of $P_{g_1}, \ldots, P_{g_{d^2}}$ the
following vectors:
\[
\vec{\lambda}\otimes (+\cdots +) \otimes w_{g_1}\,,\quad
\vec{\lambda}\otimes (+\cdots +) \otimes w_{g_2}\,,\quad\ldots\,,\quad
\vec{\lambda}\otimes (+\cdots +) \otimes w_{g_{d^2}}\,,
\]
where $(+\cdots +)$ is a vector of length $d^2$. Whenever $g_i$, $g_j$
are not both equal to the identity $g_1$ all these vectors are
orthogonal to each other since the columns of the character table are
orthogonal. This shows that all rows of the matrices $P_{g_2},\ldots,
P_{g_{d^2}}$ are orthogonal and the matrices satisfy the Schur
condition $P_g \circ P_h = P_{gh}$.}

\paragraph{Phase matrices from Hamming codes}

In the following we show how the known constructions of sign matrices
can be reproduced with well known families of orthogonal arrays. What
is more, we show that the class of orthogonal arrays used for this
construction are based on Hamming codes \cite{MS:77,Blahut:2003}. They
can be used to construct phase matrices for higher-dimensional systems
in case the dimension $d$ of the qudits is a power of a prime.

First, we briefly recall some basic facts about error-correcting codes
\cite{MS:77,Blahut:2003} since they will feature in the subsequent
constructions of orthogonal arrays. A linear code over the finite
field $\F_q$ is a $k$-dimensional subspace of the vector space
$\F_q^n$. The metric on the space $\F_q^n$ is called the Hamming
weight. For $x = (x_1, \ldots, x_n) \in \F_q^n$ we have that ${\rm
wt}(x) := |\{ i\in \{1, \ldots, n\} : x_i \not=0\}|$. The minimum
distance of a linear code $C$ is defined by $d = d_{\rm min} :=
\min{\{\rm wt}(c) : c \in C,c\neq 0\}$, where $0$ denotes the zero
vector. As a shorthand we often abbreviate this situation by saying
that $C$ is an $[n, k, d]_q$ code. We need one more definition which
is the dual code $C^\perp$ of $C$ defined by $C^\perp := \{ x \in
\F_q^n : x \cdot y = 0 \, \mbox{for all} \, y \in C\}$.

The following theorem \cite[Theorem 4.6]{HSS:99} establishes a
connection between orthogonal arrays and error-correcting codes. In
fact this is one of the most prolific constructions for OAs known.

\begin{theorem}[OAs from linear codes]\label{OAcodes}
Let $C$ be a linear $[n, k, d]_q$ code over $\F_q$. Let $d^\perp$ be
the minimum distance of the dual code $C^\perp$. Arrange the code
words of $C$ into the columns of a matrix $A \in \F_q^{n\times
q^k}$. Then $A$ is an $OA(q^k, n, q, d^\perp-1)$.
\end{theorem}

For the case of a network consisting of $n$ qubits which are governed
by a pair-interaction Hamiltonian we can construct decoupling schemes
using $N$ pulses from an $OA(N, n, $4$, 2)$. Hence, in order to apply
Theorem \ref{OAcodes} we have to find a code $C$ of linear codes over
$\F_4$ for which the parameters are $[n, k, d]$ and for which the
minimum distance $d^\perp$ of the dual code is at least $3$.

Let $q$ be a prime power and let $m\in \N$. Then the Hamming code
$H_{q,m}$ of length $n=(q^m-1)/(q-1)$ is a single-error correcting
code over the field $\F_q$ with parameters $[n,n-m,3]_q$.  The dual
code $H^\perp_{q,m}$ of the Hamming code $[n,n-m,3]_q$ has parameters
$[n,m,q^{m-1}]$. By specializing $q=4$ and by using Theorem
\ref{OAcodes} for $H^\perp_{4,m}$ we therefore obtain orthogonal
arrays with parameters $OA(N, n, 4, 2)$, where $n=(4^m-1)/3$ and $N =
4^m$ for any choice of $m \in \N$. The alphabet set is in this case
the finite field $\F_4$ of four elements.

The procedure to obtain a decoupling scheme for a network of $n$
qubits, where $n$ is an arbitrary natural number, i.\,e., not
necessarily of the form $n=(4^m-1)/3$ is as follows: first let $m\in
\N$ be the unique integer such that $n \leq \frac{4^m-1}{3} \leq
4n$. Then construct the orthogonal array with parameters $OA(4^m,
(4^m-1)/3,4,2)$. The columns of this OA are code words of
$H_{4,m}^\perp \subseteq \F_4^{(4^m-1)/3}$. We can now obtain a triple
of sign matrices $S_x$, $S_y$, and $S_z$ by using the substitution
rules in Theorem~\ref{th:phaseMatricesfromOAs}. This leads to the same
sign matrices as the ones constructed in \cite{reversal} by a direct
construction and in \cite{finite} using spreads in the geometry
$\F_2^{2m}$.

In case the dimension is an arbitrary power of a prime $d = p^r$, we
use the Hamming code $[n,n-m,3]_{d^2}$ to obtain an $OA(N, n, d^2,
2)$, where $n=(q^m-1)/(q-1)$ and $N = q^m$. By Theorem
\ref{th:phaseMatricesfromOAs} from this orthogonal array a collection
of phase matrices can be constructed which satisfy the orthogonality
conditions and are compatible with respect to taking Schur products.

\subsection{Orthogonal arrays from phase matrices}

In this section we provide a converse to the previous section by
showing that orthogonal arrays of strength two can be constructed from
phase matrices satisfying the decoupling conditions in
eqs.~(\ref{eq:removeLocal}) and (\ref{eq:removeBilinear}). To do this
we need the following lemma which gives a criterion in terms of group
characters to decide whether an element of the group ring is an
equally-weighted sum of all group elements. This allows to check
whether a matrix is an orthogonal array.

\begin{lemma}\label{lemma:equallyoften}
Let $G$ be an abelian group of order $|G|$. Denote by
$\chi_1,\chi_2,\ldots,\chi_{|G|}$ all irreducible characters of $G$,
where $\chi_1$ is the trivial character (i.\,e. $\chi_1(h)=1$ for all
$h\in G$). Let $v$ be an arbitrary element of the group ring $\C[G]$,
i.\,e., $v$ is a formal sum of (weighted) group elements
\begin{equation}\label{eq:elementGroupRing}
v:=\sum_{g\in G}\mu_g g\,,\quad\mu_g\in\C\,.
\end{equation}
If $\chi_i(v)=0$ for all $i=2,\ldots,|G|$ then we have
$
v=\frac{\mu}{|G|} \sum_{g\in G} g\,,
$
where $\mu:=\chi_1(v)=\sum_{g\in G} \mu_g$.
\end{lemma}

A proof of this lemma is given in Appendix \ref{appendixB}. We are now
ready to state the main result of this section.

\begin{theorem}\label{th:phaseOrthArray}
Let $G$ be a finite abelian group and let $[P_g]_{g\in G}$ be a
collection of phase matrices of size $n\times N$ which are compatible
with respect to taking Schur products and satisfy the orthogonality
relations. Then these phase matrices define an orthogonal array
$OA(N,n,|G|,2)$.
\end{theorem}
\proof{For fixed $k=1,\ldots,n$ and $j=1,\ldots,N$ each vector
$v_{k,j}:=[P_h;k,j]_{h\in G}$ is a row the character table of
$G$. Therefore, it determines uniquely $g\in G$ such that the entries
of $v_{k,j}$ are $\chi_g(h)$, i.\,e., the values of the irreducible
character corresponding to $g$ applied to $h\in G$. We denote the so
defined group element by $g_{k,j}$. Let $M=(g_{k,j})$ be the $n\times
N$ matrix with entries $g_{k,j}$. We will show that $M$ is an
orthogonal array $OA(N,n,|G|,2)$.

Pick any two rows $(g_{kj})$ and $(g_{\ell j})$ of $M$. We define
an element of the group ring $\C[G\times G]$ as the formal sum
\[
r_{k \ell}:=\sum_{j=1}^N (g_{kj},g_{\ell j})\,.
\]
To abbreviate the notation we denote by $\chi_{g,g'}$ the irreducible
character of $G\times G$ corresponding to the element $(g,g')$.  The
decoupling conditions given in eqs.~(\ref{eq:removeLocal}) and
(\ref{eq:removeBilinear}) are equivalent to
\[
\chi_{g,g'}(r_{k \ell})=0
\]
for all $(g,g')\neq (e,e)$. By Lemma~\ref{lemma:equallyoften} this is
equivalent to the case that all elements of $G\times G$ appear equally
often in the sum $r_{k \ell}$. This shows that $M$ is an orthogonal
array $OA(N,n,|G|,2)$ of strength $t=2$ over $G$.}

Theorems \ref{th:phaseMatricesfromOAs} and \ref{th:phaseOrthArray}
shows that phase matrices can be used to define an orthogonal array of
strength two and vice versa. Based on the above lemma we give an
alternative characterization of orthogonal arrays of arbitrary
strength which is a generalization of \cite[Theorem
3.30]{HSS:99}. This theorem implies that the entries of the array can
be replaced by complex numbers such that the resulting matrix is
orthogonal with respect to the usual inner product for strength
$t=2$. Recall that for elements of the Cartesian product $v \in G^n$
the Hamming weight ${\rm wt}(v)$ is defined by the number of
components which are different from the identity.

\begin{theorem}[Conditions for a matrix to be an OA] 
Let $G$ be a finite abelian group and let $A$ be a matrix of size $n
\times N$ with entries from $G$. Then $A$ is an orthogonal array
$OA(N, n, |G|, t)$ if and only if
\begin{equation}\label{oaChar}
\sum_{j=1}^N
\prod_{i=1}^n \chi(A_{i,j} v_i) = 0
\end{equation}
holds for all nontrivial characters $\chi \in {\rm Irr}(G)$ and for
all $v \in G^n$ of Hamming weight ${\rm wt}(v) \leq t$.
\end{theorem}
\proof{Suppose $A$ is an orthogonal array $OA(N,n,|G|,t)$ and let
$v\in G^n$ be a fixed element of weight ${\rm wt}(v) \leq t$. Let
$\vartheta$ be a non-trivial character of the $t$-fold direct product
$G^t$. Denote by $v_t \in G^t$ the vector containing the components of
$v$ which are different from the identity. We obtain that $\sum_{g \in
G^t} \vartheta(g v_t) = \sum_{g \in G^t} \vartheta(g) =
0$ and the statement follows from the fact that the characters of
$G^t$ are given by products of $t$ characters of $G$.

Conversely, assume that we are given a matrix $A$ such that
eq.~(\ref{oaChar}) holds for all non-trivial characters and all $v \in
G^n$ of weight ${\rm wt}(v) \leq t$. In particular, this means that
eq.~(\ref{oaChar}) is satisfied for the vector $w_e$ all components of
which are equal to the identity $e \in G$. Again, fix a $t$-subset $T$
of the rows of $A$. We have that for all non-trivial characters
$\vartheta$ of $G^t$ the identity $\sum_{j=1}^N \vartheta(g_j w_e) =
0$ holds, where the elements $g_j \in G^t$ are obtained by 
selecting the $j$th column of $A$, followed by selecting the
components corresponding to $T$, and finally to consider the
element as being an element of $G^t$. Now, we can apply Lemma
\ref{lemma:equallyoften} to obtain that the list $\left[ g_j : j = 1,
\ldots, N \right]$ has to contain all elements $G^t$ and that each
element has to occur the same number of times. This shows that $A$ is
an $OA(N,n,|G|,t)$.}

%
%

\subsection{A new characterization of orthogonal arrays}

In Section~\ref{oa} we have seen that orthogonal arrays of strength
$t$ can be used to construct decoupling schemes for $t$-local
Hamiltonians. In order to establish a converse result we need some
additional conditions on the class of schemes considered: (i) the
schemes have to be regular (see Definition~\ref{def:decoupling}) and
moreover we will assume that (ii) the pulses are actually taken from a
fixed set of unitaries which in addition will be assumed to form a
unitary error basis. We begin by stating some standard concepts from
quantum information theory which will be used in the proof. Recall
that the {\em Shannon entropy} \index{Shannon entropy} \index{entropy}
is defined by the equation
\[
H(p_1,\ldots,p_M)=-\sum_{j=1}^M p_j \log_2 p_j\,.
\]
Shannon entropy measures the disorder of probability distributions. If
$p_j=1$ for some $j$, then the entropy is zero. The entropy takes its
maximum value $\log_2 M$ for the uniform distribution. The notion of
entropy extends to density operators, and is usually called {\em von
Neumann entropy}. Let $\rho$ be an arbitrary density operator on
$\C^d$. Then the spectral decomposition $\rho = \sum_{j=1}^M \lambda_j
\ket{\Psi_j}\bra{\Psi_j}$ is such that the eigenvalues
$\lambda_1,\ldots,\lambda_M$ form a probability distribution and the
eigenvectors $\ket{\Psi_1},\ldots,\ket{\Psi_M}$ form an orthogonal
basis of $\C^M$. The von Neumann entropy $S(\rho)$ of $\rho$ is
defined by the equation
\[
S(\rho)=-\sum_{j=1}^M \lambda_j \log_2\lambda_j\,.
\]
The von Neumann entropy takes its minimal value $0$ on pure states,
i.\,e., for $\rho=\ket{\Psi}\bra{\Psi}$, and its maximal value $\log_2 M$
for the maximally mixed state $\rho=\onemat/M$. Let
$U_1,\ldots,U_N\in\C^{M\times M}$ be arbitrary unitary matrices,
$p_1,\ldots,p_N$ a probability distribution, and $\ket{\Psi}$ a state
of $\C^M$. We have the following inequality (see \cite{NC:00}, p.~518)
\begin{equation}\label{eq:entropyIneq}
S\left(\sum_{j=1}^N p_j U_j^\dagger\ket{\Psi}\bra{\Psi} U_j\right) \le
H(p_1,\ldots,p_N) \le \log_2 N\,.
\end{equation}
The following theorem shows that if a regular decoupling scheme can
switch off an arbitrary $t$-local Hamiltonian, then the tensor
products of the operations performed on an arbitrary $t$-tuple of
qudits must form a unitary error basis for this subsystem.

\begin{theorem}[Equivalence of decoupling schemes and OAs]
\label{decouple}${}$\\
Let $D$ be a regular decoupling scheme that uses elements of a unitary
error basis ${\cal E}:=\{U_1,\ldots,U_{d^2}\}$ as control operations,
acts on $n$ qudits and consists of $N$ time-slots. Denote by
$U_{m_{1j}}\otimes U_{m_{2j}}\otimes\cdots\otimes U_{m_{nj}}$ the
local operation that is performed on the qudits in time-slot
$j=1,\ldots,N$, i.\,e., the indices $m_{kj}\in\{1,\ldots,d^2\}$
determine which elements of ${\cal E}$ are applied to the qudits in
the time-slots. The scheme $D$ can be used to decouple any $t$-body
Hamiltonian if and only if the matrix $M=(m_{kj})$ where
$k=1,\ldots,n$ and $j=1,\ldots,N$ is an orthogonal array
$OA(N,n,d^2,t)$ of strength $t$.
\end{theorem}
\proof{Consider a fixed $t$-tuple $(k_1,\ldots,k_t)$ with different entries
from $\{1,\ldots,n\}$. Let
\[
H_{k_1\ldots, k_t} :=
\sum_{s=1}^t 
\sum_{(\ell_1,\ldots, \ell_s)}
\sum_{\alpha_1\ldots\alpha_s}
J_{(\ell_1,\ldots, \ell_s);\alpha_1\ldots\alpha_s}
\sigma_{\alpha_1}^{(\ell_1)}\sigma_{\alpha_2}^{(\ell_2)}\cdots
\sigma_{\alpha_s}^{(\ell_s)}\,,
\]
where $(\ell_1,\ldots,\ell_s)$ runs over $s$-tuples with different
entries from $\{k_1,\ldots,k_t\}$. We say that the operator
$H_{k_1,\ldots,k_t}$ is the restriction of the $t$-body Hamiltonian
$H$ to the qudits $k_1,\ldots,k_t$. We denote by
$\hat{H}_{k_1,\ldots,k_t}$ the corresponding operator acting on
$(\C^{d})^{\otimes t}$.

Note that for every traceless Hermitian operator $X$ acting on
$(\C^d)^{\otimes t}$ there is a $t$-local Hamiltonian $H$ such that
its restriction $H_{k_1,\ldots,k_t}$ to the qudits $k_1,\ldots,k_t$ is
given by the embedding $X^{(k_1,\ldots,k_t)}$ of $X$ to
$(\C^d)^{\otimes n}$ according to the tuple $(k_1,\ldots,k_t)$. Let
$T_D$ be the operator
\[ 
T_D : H \mapsto \sum_{i=1}^N p_j 
(U_{m_{1j}} \otimes U_{m_{2j}} \otimes \cdots \otimes U_{m_{nj}})^\dagger 
H
(U_{m_{1j}} \otimes U_{m_{2j}} \otimes \cdots \otimes U_{m_{nj}})
\]
Define the weight $w_{i_1,\ldots,i_t}$ of each tuple
$(i_1,\ldots,i_t)\in \{1,\ldots,d^2\}^t$ to the sum of all $p_j$'s
with $(m_{k_1,j},\ldots,m_{k_s,j})=(i_1,\ldots,i_t)$. Now suppose that
$T_D(H) = \zeromat$ for all $t$-body Hamiltonians. Consequently, we
have that $T_D(H_{k_1,\ldots,k_t})=\zeromat$ for all restrictions to
$t$-tuples. But this implies that the weights for all ${\cal A}^t$
must be equal. This is seen as follows: the equality
\begin{eqnarray*}
T_D(H_{k_1\ldots k_t}) 
& = &
\left[ 
\sum_{i_1,\ldots,i_t=1}^{d^2}
w_{i_1\ldots i_t} 
(U_{i_1}\otimes\cdots\otimes U_{i_t})^\dagger \, 
\hat{H}_{k_1\ldots k_t} \,
(U_{i_1}\otimes\cdots\otimes U_{i_t})
\right]^{(k_1,\ldots,k_t)} \\
& = &
\zeromat^{(k_1,\ldots,k_t)} = \zeromat
\end{eqnarray*}
shows that the operation defined by the sum above is a unitary
depolarizer for $(\C^d)^{\otimes t}$, i.\,e.,
\[
\sum_{i_1,\ldots,i_t=1}^{d^2}
w_{i_1\ldots i_t} 
(U_{i_1}\otimes\cdots\otimes U_{i_t})^\dagger \, 
X
(U_{i_1}\otimes\cdots\otimes U_{i_t}) = \frac{tr(X)}{d^t}\onemat 
\]
for all operators $X$ acting on $(\C^d)^{\otimes t}$. Now, we show
that the weights must be all equal. Let
$\ket{\Psi_1},\ldots,\ket{\Psi_{d^t}}$ be an orthonormal basis of
$(\C^d)^{\otimes t}$. We define a special state in the bipartite
system $(\C^d)^{\otimes t} \otimes (\C^d)^{\otimes t}$ together with
its corresponding density operator
\[
\ket{\Psi}=
\frac{1}{\sqrt{d^t}}\sum_{r=1}^{d^t} \ket{\Psi_r}\otimes\ket{\Psi_r}\,,\quad
\rho =
\frac{1}{d^t}\sum_{r,s=1}^{d^t}
\ket{\Psi_r}\bra{\Psi_s}\otimes\ket{\Psi_r}\bra{\Psi_s}\,.
\]
We use the fact that $T_D$ is a unitary depolarizer to show that show
that all weights are equal
\begin{eqnarray*}
\onemat\otimes T_D\, (\rho) & = & 
\sum_{i_1\ldots i_t\in {\cal A}^t} w_{i_1\ldots i_t} 
(\onemat_{d^t} \otimes
U_{i_1}\otimes\cdots\otimes U_{i_t})^\dagger
\rho\,
(\onemat_{d^t} \otimes
U_{i_1}\otimes\cdots\otimes U_{i_t}) \\
& = & 
\frac{1}{d^t}\sum_{r,s=1}^{d^t}
\ket{\Psi_r}\bra{\Psi_s}\otimes \\
& &
\quad
\sum_{i_1\ldots i_t\in {\cal A}^t} w_{i_1\ldots i_t}
(U_{i_1}\otimes\cdots U_{i_t})^\dagger
\,\ket{\Psi_r}\bra{\Psi_s}\,
(U_{i_1}\otimes\cdots U_{i_t}) \\ 
& = &
\frac{1}{d}\sum_{r=1}^d \ket{\Psi_r}\bra{\Psi_r}\otimes
\onemat_{d^t}/d^t \\
& = & \onemat_{d^t}/d^t\otimes\onemat_{d^t}/d^t = 
\onemat_{d^{2t}}/d^{2t}\,.
\end{eqnarray*}
It follows from the above equation that we need at least $d^{2t}$
different unitaries are necessary since the rank of each pure state
$(\onemat\otimes U^\dagger)\ket{\Psi}\bra{\Psi}(\onemat\otimes U)$ is
one and since they have to sum up to $d^{2t}$ (the rank of the maximally
mixed state).

Since we use exactly $d^{2t}$ different unitaries (tensor products of
elements of the unitary error basis $\cE$) as control operations all
weights $w_{i_1\ldots i_t}$ must be equal due to the
inequality~(\ref{eq:entropyIneq}). Now together with the fact that for
regular schemes all time-slots have equal length we conclude that
$U_{i_1}\otimes\cdots\otimes U_{i_t}$ must appear with the same
multiplicity. Therefore, by considering all $t$-tuples
$k_1,\ldots,k_t$ of $t$ qudits we see that the decoupling $D$ scheme
must correspond to an orthogonal array $OA(N,n)$ with $d^2$ levels and
strength $t$.}

%
%

\section{Conclusions}

We have shown the equivalence between two constructions for decoupling
schemes and selective coupling schemes in networks of qubits. One
construction is based on triples of sign matrices which are closed
under taking entry-wise products, while the other construction is
based on orthogonal arrays of strength two over an alphabet of size
four.

The construction using orthogonal arrays can be generalized to systems
where the nodes have higher dimensions. Also the case where the system
Hamiltonian has higher couplings can be dealt with by using orthogonal
arrays: the coupling order directly translates into the strength of
the orthogonal array. A special case arises when the local pulses
which are applied in each time-slot are actually elements of a nice
error basis for an abelian group. We have shown that in case of
equidistant interval lengths (after refinement) this leads to a class
of schemes which are equivalent to orthogonal arrays. In addition we
have presented a construction of schemes for decoupling and selective
coupling which can be constructed by using Hamming codes.

Moreover, we have shown that the construction of this particular class
of decoupling and coupling schemes can be reduced to questions about
the existence of these combinatorial arrays. While several
constructions for orthogonal arrays are known, there still remain some
open problems such as the case where the dimensions of the nodes could
be different. Another important problem is to devise schemes for a
situation where the given Hamiltonian is of a particular form, i.\,e.,
where not all interactions are present or can be assumed to be very
weak for a large number of pairs. In this case a combination of graph
theoretical techniques and the methods described in this paper can be
developed.

\section*{Acknowledgments}
It is a pleasure to thank Thomas Beth and Dominik Janzing for useful
discussions. M.~R.\ has been supported in part by NSA and ARDA under
the ARDA Quantum Computing Program. He also acknowledges support by
CFI, ORDCF, and MITACS. P.~W.\ has been supported by the BMBF-project
01/BB01B and the National Science Foundation under grant EIA-0086038
through the Institute for Quantum Information at the California
Institute of Technology.

%
%

%
%

\begin{appendix}

\section{Appendix: Characters of abelian groups}
\label{characters}

In this appendix we collect some basic facts of the representation
theory of finite groups which are needed in the paper. Recall that
$\GL(n,\C)$ denotes the group of invertible $n\times n$ matrices with
entries in $\C$. Let $G$ be a finite group.  A {\em representation} of
$G$ over $\C$ (see also \cite{Isaacs:76,Huppert:83}) is a homomorphism
$\rho$ from $G$ to $\GL(n,\C)$, for some $n\in \N$. The {\em degree}
of $\rho$ is given by $n$. The representation $\rho$ is called {\em
irreducible} if there are no invariant subspaces under the action of
the matrices $\{\rho(g)\}_{g\in G}$ apart from the trivial subspaces
$\{0\}$ and $\C^n$. With each $n\times n$ matrix $\rho(g)$ we
associate the complex number given by the trace of $\rho(g)$, and call
this number $\chi(g)$. The function $\chi : G\rightarrow \C$ is called
the character of the representation $\rho$. A character is called
irreducible if the corresponding representation is irreducible.  We
need the following theorem on abelian groups \cite[Chap.~V. \S
6]{Huppert:83}.
\begin{theorem}[Characters of abelian groups]\label{chgroup}
Let $G$ be a finite abelian group of order $|G|$. Then every
irreducible representation $\rho$ of $G$ has degree $1$, i.\,e., we
have that $\rho:G\rightarrow\C^{\times}$ is a homomorphism which maps
$G$ to scalars. Furthermore, the number of different irreducible
representations (irreducible characters) of $G$ is given by $|G|$ and
the characters form a group $\hat{G}={\rm Hom}(G,\C^{\times})$ under
pointwise multiplication. Hence, we have that
\[
\chi\tilde{\chi}(h)=\chi(h)\tilde{\chi}(h)
\]
for all irreducible characters $\chi,\tilde{\chi}$ and $h\in
G$. Moreover, the character group $\hat{G}$ is isomorphic to
$G$. Thus, we can label the characters of $G$ by the elements of $G$
using an isomorphism which maps $h \mapsto \chi_h$ for all $h \in G$.
\end{theorem}

\section{Appendix: Proof of Lemmas \ref{lemma:character} and
\ref{lemma:equallyoften}}
\label{appendixB}

{\bf Lemma \ref{lemma:character}:} {\it Let $\cE:=\{U_g\mid g\in G\}$
be a nice error basis with an abelian index group $G$. Then the 
matrix ${\cal X} = \left( \chi(g,h) \right)_{g,h\in G}$ is the
character table of the group $G$.}

\proof{Let $\alpha$ be the factor system corresponding to the nice
error basis $\cE$ with abelian index group $G$. We prove that ${\cal
X}$ is a character table by showing that the rows of ${\cal X}$ form a
group under pointwise multiplication that is isomorphic to $G$ (see
Theorem~\ref{chgroup} in Appendix).  We first show that
\begin{equation}
\chi(g,h)=\frac{\alpha(h,g)}{\alpha(g,h)}\,.
\end{equation}
We have that
\begin{eqnarray}
U_g U_h & = & \alpha(g,h) U_{gh}, \label{eq:gh}\\ 
U_h U_g & = & \alpha(h,g) U_{hg} = \alpha(h,g) U_{gh}\,.\label{eq:hg}
\end{eqnarray}
By multiplying eq.~(\ref{eq:hg}) by $U_g^\dagger$ from the left and
using eq.~(\ref{eq:gh}) we obtain
\begin{eqnarray*}
U_g^\dagger U_h U_g & = & \alpha(h,g) U_g^\dagger U_{gh} \\
& = & \frac{\alpha(h,g)}{\alpha(g,h)} U_g^\dagger U_g U_h \\
& = & \frac{\alpha(h,g)}{\alpha(g,h)} U_h\,.
\end{eqnarray*}

We now prove that the rows of ${\cal X}$ form a group under pointwise
multiplication that is isomorphic to $G$. Let $g,\tilde{g}$ be
arbitrary elements of $G$. Note that we have
$\overline{\alpha(\tilde{g}^{-1},g)}\alpha(\tilde{g}^{-1},g)=1$
(otherwise the matrix
$U_{\tilde{g}^{-1}}U_g=\alpha(\tilde{g}^{-1},g)U_{\tilde{g}^-1 g}$
would not be unitary). The group property is verified by
\begin{eqnarray*}
\chi(g,h)\chi(\tilde{g}^{-1},h) U_h & = &
U_g^\dagger U_{\tilde{g}^{-1}}^\dagger U_h U_{\tilde{g}^{-1}} U_g \\
& = &
\overline{\alpha(\tilde{g}^{-1},g)}\alpha(\tilde{g}^{-1},g)\,
U_{g\tilde{g}^{-1}}^\dagger U_h U_{g\tilde{g}^{-1}} \\
& = &
U_{g\tilde{g}^{-1}}^\dagger U_h U_{g\tilde{g}^{-1}} \\
& = &
\chi(g\tilde{g}^{-1},h) U_h
\end{eqnarray*}
for all $h\in G$. 

The rows of ${\cal X}$ form a group that is isomorphic to $G$ (and not
only to a proper subgroup of $G$) since there is a bijection between
the rows of ${\cal X}$ and the elements of $G$. This is seen as
follows. Assume that there are $g\neq\tilde{g}$ such that
$\chi(g,h)=\chi(\tilde{g},h)$ for all $h\in G$. This is equivalent to
$U_g^\dagger U_h U_g = U_{\tilde{g}}^\dagger U_h U_{\tilde{g}}$.  Set
$U=U_{\tilde{g}} U_g^\dagger$. Then we have $U M=M U$ for all
$M\in\C^{d\times d}$ since the matrices $U_h$ form a basis of
$\C^{d\times d}$. Therefore $U$ must be a multiple of the identity
matrix. Due to the properties of a nice error basis this is only
possible for $g=\tilde{g}$. This proves that there is a bijection
between the group elements of $G$ and the rows of ${\cal X}$.}

\noindent
{\bf Lemma \ref{lemma:equallyoften}:} {\it Let $G$ be an abelian group
of order $|G|$. Denote by $\chi_1,\chi_2,\ldots,\chi_{|G|}$ all
irreducible characters of $G$, where $\chi_1$ is the trivial character
(i.\,e.\ $\chi_1(h)=1$ for all $h\in G$). Let $v$ be an arbitrary
element of the group ring $\C[G]$, i.\,e., $v$ is a formal sum of
(weighted) group elements
\begin{equation}
v:=\sum_{g\in G}\mu_g g\,,\quad\mu_g\in\C\,.
\end{equation}
If $\chi_i(v)=0$ for all $i=2,\ldots,|G|$ then we have
$
v=\frac{\mu}{|G|} \sum_{g\in G} g\,,
$
where $\mu:=\chi_1(v)=\sum_{g\in G} \mu_g$.}

\proof{Let
$G:=\{g_1,\ldots,g_{|G|}\}$ be an arbitrary ordering of the group
elements, where $g_1$ is the identity element of $G$. Denote by ${\cal X}$
the (normalized) character table of $G$, i.\,e.,
\begin{equation}
{\cal X}_{ij}:=|G|^{-1/2}\,\chi_i(g_j)
\end{equation}
for $i,j=1,\ldots,|G|$. Recall that the (normalized) character table
${\cal X}$ is a unitary matrix and has the following form
\cite{Huppert:83,Isaacs:76}
\begin{equation}\label{eq:formCharacterTable}
{\cal X}=\frac{1}{|G|^{1/2}} 
\left(
\begin{array}{ccc}
1      & \cdots & 1 \\
\vdots & *      &   \\
1      &        &
\end{array}
\right).
\end{equation}
The conditions given in the lemma can now be expressed as
\[
|G|^{1/2} {\cal X} 
\left(
\begin{array}{c}\mu_1\\\mu_2\\\vdots\\\mu_{|G|}\end{array}
\right)
=
\left(
\begin{array}{c}\mu\\ 0\\\vdots\\ 0\end{array}
\right)\,.
\]
Multiplying by the inverse ${\cal X}^{-1}$ we obtain
\[
(\mu_1,\mu_2,\ldots,\mu_{|G|})^T=
\frac{\mu}{|G|} (1,1,\ldots,1)^T\,.
\]
due to the special form in eq.~(\ref{eq:formCharacterTable}). This
show that all coefficients $\mu_g$ in eq.~(\ref{eq:elementGroupRing})
are equal to $\mu/|G|$.}

\end{appendix}

\end{document}